\documentclass[12pt]{article}
\usepackage{graphicx}
\usepackage{amsfonts}
\usepackage{amssymb,amsmath}
\usepackage{color}
\usepackage[colorlinks=true,linkcolor=blue,citecolor=blue]{hyperref}
\usepackage{bm}

\newcommand{\be}{\begin{eqnarray}}
\newcommand{\ee}{\end{eqnarray}}

\begin{document}

\begin{titlepage}

\begin{flushright}
RESCEU-8/18
\end{flushright}

\begin{center}

{\Large \bf 
Long-term dynamics of cosmological axion strings
}

\vskip .45in

{
Masahiro Kawasaki,$^{a,b}$
Toyokazu Sekiguchi,$^{c}$
Masahide Yamaguchi$^{d}$ \\
and 
Jun'ichi Yokoyama$^{b,c,e}$ 
}

\vskip .45in

{\it 
$^a$ ICRR, University of Tokyo, Kashiwa, 277-8582, Japan\\
$^b$ Kavli Institute for the Physics and Mathematics of the Universe (Kavli IPMU), WPI, UTIAS, The University of Tokyo, Kashiwa, Chiba 277-8568, Japan\\
$^c$ Research Center for the Early Universe (RESCEU), Graduate School of Science, The University of Tokyo, Tokyo 113-0033, Japan\\
$^d$ Department of Physics, Tokyo Institute of Technology, 2-12-1 Ookayama, Meguro-ku, Tokyo 152-8551, Japan\\
$^e$ Department of Physics, Graduate School of Science, The University of Tokyo, Tokyo 113-0033, Japan\\
}

\end{center}

\vskip .4in

\begin{abstract}
We present results of new field-theoretic simulation of cosmological axion strings,
which are eight
times longer than previous ones.
We have upgraded our simulation of physical strings in Hiramatsu {\it et al.} (2011) in terms of the number of grids 
as well as the suite of analysis methods. These improvements enable us to monitor a variety of quantities characterizing 
the dynamics of the physical string network for the longest term ever. 
Our extended simulations have revealed that global strings do not evolve according to the scaling solution but its scaling parameter, 
or the number of long strings per horizon, increases logarithmically in time.
In addition, we have also found that the scaling parameter shows nontrivial dependence on the breaking scale of the Peccei-Quinn symmetry.
\end{abstract}

\end{titlepage}

\setcounter{page}{1}

\section{Introduction} 
\label{sec:introduction}

The axion is one of the best motivated particles beyond the Standard
Model \cite{Weinberg:1977ma,Wilczek:1977pj}.  Its existence is
inevitably derived from the Peccei-Quinn (PQ) mechanism
\cite{Peccei:1977hh}, which dynamically solves the strong CP problem in
quantum chromodynamics (QCD). 
Concurrently, it can account for the cold dark matter (CDM) in the
observed Universe and offer rich phenomenology in the cosmological
context (see e.g. \cite{Kawasaki:2013ae,Marsh:2015xka} for recent
review).  There are a few major production mechanisms of the axion CDM in
the early Universe, including the misalignment and topological
defects. A number of experiments are going on aiming at the detection of
the CDM axion \cite{Sikivie:1983ip} (for review we refer to
\cite{Battesti:2007um,Graham:2015ouw,Irastorza:2018dyq,Tanabashi:2018???}).

Formation of topological defects associated to the broken global PQ U(1)
symmetry takes place if the symmetry breaking occurs after the
observable Universe exits horizon during inflation.
At first, axion strings form following the PQ phase transition.  As the
cosmological network of the strings evolves, its energy is released in
the form of axion radiation.  Later on, the QCD phase transition takes
place and domain wall (DW) stretches out between the axion strings.  If
the number of DWs attached to each string is unity, the network of
string-DW is unstable, so that it disappears soon after the QCD phase
transition leaving additional contribution to the axion CDM
\cite{Kawasaki:2014sqa}.  Otherwise, DWs are stable and dominate the
Universe, which spoils successful subsequent cosmology (see also
\cite{Hiramatsu:2012sc}). 

In this paper, we will focus on the network of axion strings in between
the two phase transitions. This subject has already been studied by many
different people for decades
\cite{Davis:1985pt,Davis:1986xc,Harari:1987ht,Davis:1989nj,Dabholkar:1989ju,Hagmann:1990mj,Battye:1993jv,Battye:1994au,Yamaguchi:1998gx,Yamaguchi:1999yp,Yamaguchi:1999dy,Hagmann:2000ja,Yamaguchi:2002zv,Yamaguchi:2002sh}.
However, the current picture of the string dynamics could not be
developed straightforwardly.  Notably, there had been long-standing
controversies concerning, for instance, the number of strings per
horizon and energy spectrum of radiated axions
\cite{Davis:1985pt,Davis:1986xc,Harari:1987ht,Davis:1989nj,Dabholkar:1989ju,Hagmann:1990mj,Battye:1993jv,Battye:1994au,Yamaguchi:1998gx,Hagmann:2000ja}. Those
controversies were settled principally with the aid of field theoretic
simulation of the strings based on the first principles
\cite{Yamaguchi:1998gx,Yamaguchi:1999yp,Yamaguchi:1999dy,Yamaguchi:2002zv,Yamaguchi:2002sh,Hiramatsu:2010yu,Hiramatsu:2012gg}. Nonetheless,
as we shall reveal its novel dynamics in this Letter, axion strings seem
still concealing its nature from us.  We need to pursue better
understanding of dynamics of global strings in order for accurate
prediction of the axion CDM abundance.

The primary purpose of this paper is to update the result of our
previous simulation~\cite{Hiramatsu:2010yu}.  In that paper, we
performed field theoretic simulation of cosmological axion strings and
estimated the relic abundance of axion radiated from the strings. Our
simulation is improved mainly in two respects.  Exploiting massively
parallel computation on computer clusters, the number of grids is
increased from $512^3$ to $4096^3$, which enhances the simulation time
by a factor of eight.
This allows us to examine long-term behavior of the string network.
We also improved our analysis method. In \cite{Hiramatsu:2010yu}, we introduced a novel string identification method 
and statistical reconstruction of the axion energy spectrum. In the present analysis, we incorporated estimation of velocity of strings and loop identification of strings
in the suite of analysis. This allows us to monitor a variety of quantities characterizing the string network and examine its dynamics in detail.

This paper is organized as follows. In the next section, we will first
describe the Lagrangian of the PQ scalar we simulate as well as the
essence of our numerical simulation and analysis. Then in Section
\ref{sec:results} we will present relevant results from our
simulation.  In particular, we pay particular attention to the long-term
behavior of the string network, that is found in simulation of physical
strings for the first time.  The final section will be devoted to
discussion.

\section{Setup and analysis methods} 
\label{sec:setup}

\subsection{Model} 
\label{sec:model}
We adopt the following Lagrangian density for the PQ 
complex scalar $\Phi$ as in \cite{Yamaguchi:2002sh}: 
\be
\mathcal L= -|\partial_\mu\Phi|^2
-V_{\rm eff}[\Phi;T], 
\ee
where the effective potential $V_{\rm eff}[\Phi;T]$ is dependent on the temperature $T$:
\be
  V_\mathrm{eff}[\Phi;T]=
  \lambda(|\Phi|^2-v^2)^2+\frac{\lambda}{3}T^2|\Phi|^2,
\ee
with $v$ being the vacuum expectation value of $\Phi$ at $T=0$. 
The QCD axion $\varphi$ is identified as the phase of $\Phi$. At low energy, $\Phi$ can be written as $\Phi= \frac{f_a}{\sqrt2} e^{i\varphi/f_a}$, where
$f_a=\sqrt2 v$ is the decay constant of the axion.
The mass of PQ field is also temperature dependent and given by $m^2(T)=\lambda(\frac{T^2}3-2v^2)$.
Given the potential, the PQ phase transition occurs at the critical temperature $T_c=\sqrt6v$. 
In what follows, quantities measured at $T=T_c$ are subscripted with $c$.

We assume a flat Universe with its line element
given by
\be
ds^2=-dt^2+a(t)^2d\bm x^2.
\ee
We normalize the scale factor at the PQ phase transition ({\it i.e.} $a_c=1$).
We assume the radiation domination and the Hubble expansion rate $H$ is given by
\be
H^2=\frac{\pi^2}{90} g_* \frac{T^4}{M_*^2}, 
\ee
where $g_*$ and $M_*$ are the relativistic degrees of freedom and the reduced Planck mass, respectively.
Note that $t\propto a^2$ and the conformal time $\tau$ is given by $t/2a$ in the radiation domination. 
In what follows, we denote a derivative with respect to $t$ and $\tau$ with a dot and a prime ({\it i.e.} $\frac{\partial f}{\partial t}=\dot f$ 
and $\frac{\partial f}{\partial \tau}=f'$), respectively.

Following \cite{Yamaguchi:1998gx}, we adopt a parameter
\be 
\zeta=\left(
\frac{45 \lambda M_*^2}{2\pi^2 g_* v^2}
\right)^{1/2}.
\ee
Granting that the physical width of the axion string $d$ is given by $d\simeq (\sqrt{\lambda}v)^{-1}$, $\zeta$ parameterizes 
the ratio of the Hubble scale to the string width at the time of the PQ phase transition.
In the following analysis, we fix $g_*=1000$\footnote{
The choice of $g_*$ is not relevant in our analysis, where we focus on slowly time-varying quantities such as the string parameter $\xi$
and the ratio of radiation momentum to the Hubble rate $\epsilon$. We chose the value of  $g_*$ so that the direct comparison with 
the previous studies~\cite{Yamaguchi:1998gx,Hiramatsu:2010yu} can be straightforward. 
} and $\lambda=1$ but vary $v$ among $v/M_*=0.005$, $0.002$ and $0.001$, 
which respectively give $\zeta=9.5$, $23.9$ and $47.8$. 

\subsection{Simulation} 
\label{sec:simulation}

We start the simulation immediately before the PQ phase transition, and initial condition of the real and imaginary parts of the 
PQ field $\Phi=(\phi_1+i\phi_2)/\sqrt2$ is generated assuming Gaussian random field with thermal distribution:
\be
\langle \phi_a(\bm x,t) \phi_b(\bm y,t) \rangle
&=&\frac{\delta_{ab}}{a^3}\int \frac{d^3k}{(2\pi)^3}
\frac1{E_{\bm k}}
\frac1{e^{E_{\bm k}/T}-1}e^{-i\bm k\cdot(\bm x-\bm y)}, \\
\langle \dot\phi_a(\bm x,t) \dot\phi_b(\bm y,t) \rangle
&=&\frac{\delta_{ab}}{a^3}\int \frac{d^3k}{(2\pi)^3}
\frac{E_{\bm k}}{e^{E_{\bm k}/T}-1}e^{-i\bm k\cdot(\bm x-\bm y)}, \\
\langle\phi_a(\bm x,t) \dot\phi_b(\bm y,t) \rangle&=&0,
\ee
where $a,b=1,2$ and $E_{\bm k}=\sqrt{\frac{k^2}{a^2}+m^2}$.

We integrate the classical equation of motion of $\Phi$ using the leapfrog method. 
For detailed description of how the equation is discretized on the lattice, 
we refer readers to the appendices of \cite{Yamaguchi:2002sh}. The number of grids is $N_g=4096^3$, 
which is 512 times as large as in the previous simulations \cite{Hiramatsu:2010yu}. 
In order to suppress boundary effects, the size of the comoving simulation box $L$ is set to twice as
large as the horizon scale at the final time (i.e. $L=\frac{2}{a_f H_f}$), where
the subscript $f$ indicates the final time. The final time $t_f$ is determined by $t_f/d\simeq N_g^{1/3}/4\simeq 10^3$ 
so that the lattice spacing does not exceed the string width at $t_f$.

\subsection{Analysis} 
\label{sec:analysis}
We perform a suite of analysis in order to extract a variety of dynamical quantities associated to the string network.
One of the essential techniques in the analysis is identification of axion strings from the discretized data of the PQ field on grids. 
We adopt the same string identification method as in \cite{Hiramatsu:2010yu}, which tells us not only the existence of a string inside a given cell, 
but also its location in the cell if it does exist. 

Furthermore, this time we introduce a novel method to identify string loops from the string points obtained from the string identification method.
This is realized by grouping those string points by the Friends-of-Friends algorithm.
We found that setting the physical linking length to $(d^2 t)^{1/3}$ works in practice. 
Once a distinct loop is identified, we can compute its circumference length. 
By accumulating them we estimate the string parameter $\xi$, which is defined by
\be
\rho_{\rm string}=\frac{\xi\mu}{ t^2},
\ee
where $\mu$ is the tension of the axion string.
The physical meaning of $\xi$ is the average number of strings per horizon volume. 

We also implant an estimation method of string speed following \cite{Yamaguchi:2002sh}.
On a point in the vicinity of a string, the string velocity can be estimated by
\be
\bm v=\frac{(\nabla \Phi\times\nabla \Phi^*)\times (\Phi'\nabla\Phi^*-\Phi'^*\nabla\Phi)}{(\nabla \Phi\times\nabla \Phi^*)^2}.
\ee
This is not exactly the same estimation method as in  \cite{Yamaguchi:2002sh}, but coincides 
when $\Phi$ vanishes on the point where we are trying to estimate the string velocity.
Sometimes we found the magnitude of $\bm v$ exceeds unity. This failure of estimation is caused 
where a string has large curvature and/or strings are colliding. 
However, the fraction of the failure is at worst five percent and 
does not significantly bias the estimate of the rms of $\bm v$ that will be presented in the next section.

On the other hand, the energy spectrum of the axion radiation is computed according to the pseudo-power spectrum estimator \cite{Hiramatsu:2010yu}.
The method consists mainly of two processes. First, we configure a window function which masks cells in proximity to string cores and
compute the energy spectrum of axion convolved with the window function. Then we reconstruct the energy spectrum
statistically by deconvolving the mode-mixing caused by the window function. 
This procedure can remove the contamination from the string cores, which 
dominates over the contribution genuinely from the axion radiation at high wave numbers.

More detailed description of our simulation and analysis will be presented in a separate paper \cite{Kawasaki}.

\section{Results} 
\label{sec:results}

\begin{figure}
\centering
\begin{tabular}{cc}
\includegraphics[scale=.8]{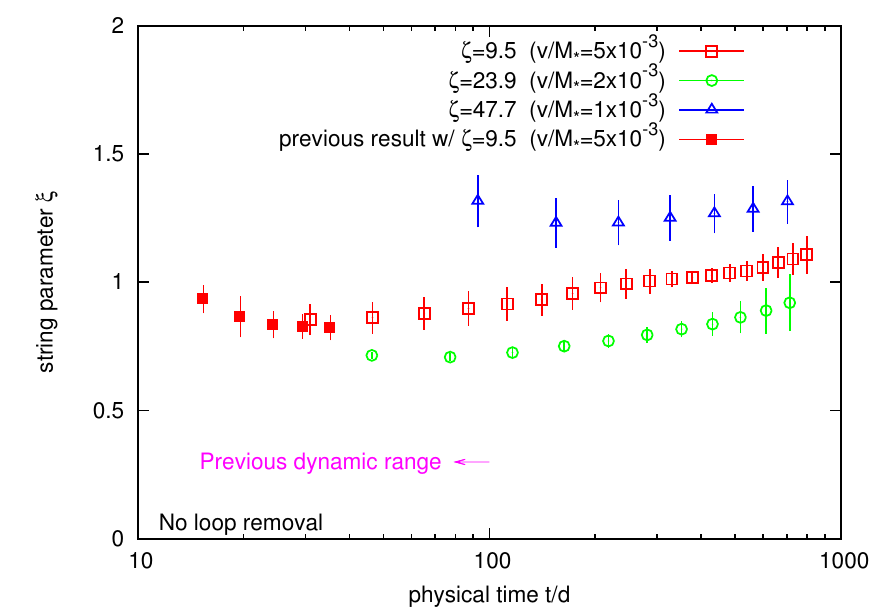} &
\includegraphics[scale=.8]{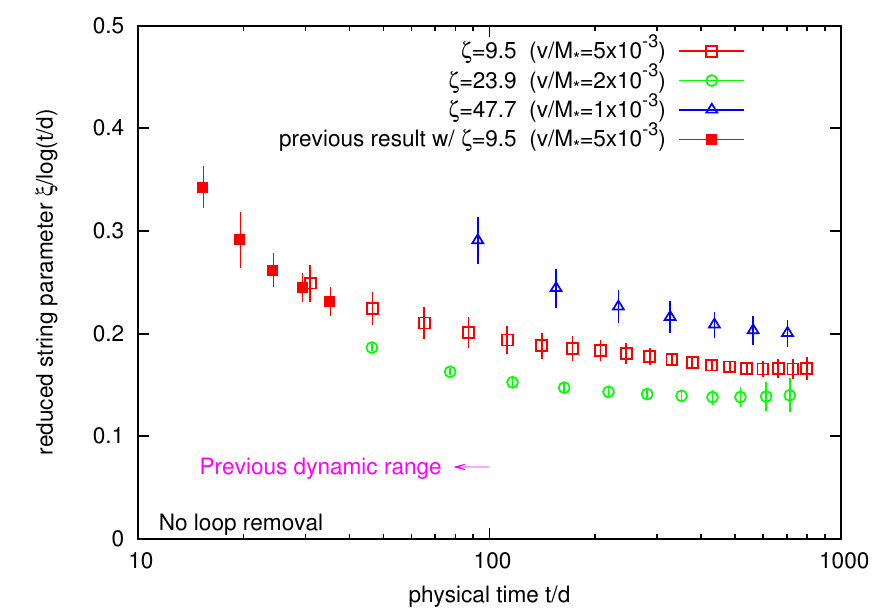}
\end{tabular}
\caption{\label{fig:xi}
Evolution of string parameter $\xi$ (left) and $\xi/\log(t/d)$ (right). Shown are the cases with three different $v/M_*=0.005$ (red open square), 
$v/M_*=0.003$ (green open circle), and $v/M_*=0.001$ (blue open circle). 
For reference, we plotted the result of simulation corresponding to our previous study \cite{Hiramatsu:2010yu} (red filled square).
Error bar shows the standard deviation among realizations. Magenta arrow shows the dynamic range that 
should have been available in the previous simulation \cite{Hiramatsu:2010yu}.}
\end{figure}

Figure~\ref{fig:xi} shows the time-evolution of the string parameter
$\xi$ with $v/M_*=0.005$ (red open square), 0.002 (green open circle),
and 0.001 (blue open circle).  
For reference, simulation result with
the same setup as in our previous study \cite{Hiramatsu:2010yu} is also
shown (red filled square), whose dynamic range is only as large as
$t_f/d\simeq40$.\footnote{
We note that our previous choice of $t_f/d=40$ is conservative compared to that expected from $N_g$ (magenta
arrow).  This makes the actual enhancement of the dynamic range more than eight and actually close to twenty.} 
 As can be read from the
figure, $\xi$ grows in time and 
the time scale of the growth can be
asymptotically characterized in 
$\ln(t/d)$.\footnote{
We note
that when we fit $\xi$ with a power function of $\log(t/d)$, that
is, $[\log(t/d)]^\alpha$ at $\log(t/d)\ge 300$, the best-fit power
indices $\alpha$ are $0.64\pm0.02$, $0.98\pm0.03$ and $0.45\pm0.02$ for $v/M_*=0.005$, 0.002, and
0.001, respectively.}  Although the magnitude of $\xi$ does not agree
among $v/M_*=0.005$, $0.002$ and $0.001$, the asymptotic
logarithmic growth is common.  Such logarithmic growth has not been
reported in the previous studies of physical strings including ours
\cite{Hiramatsu:2010yu} due to the limitation of the dynamic range.
Those simulations with such small dynamic range could trace the
evolution only up to $t/d\lesssim\mathcal O(10)$ where such
long-term dynamics has been buried in statistical uncertainties. The
larger number of grids in our simulation, which enables longer duration
of simulation, is essential in finding this novel property.

\begin{figure}
\centering
\begin{tabular}{cc}
\includegraphics[scale=.8]{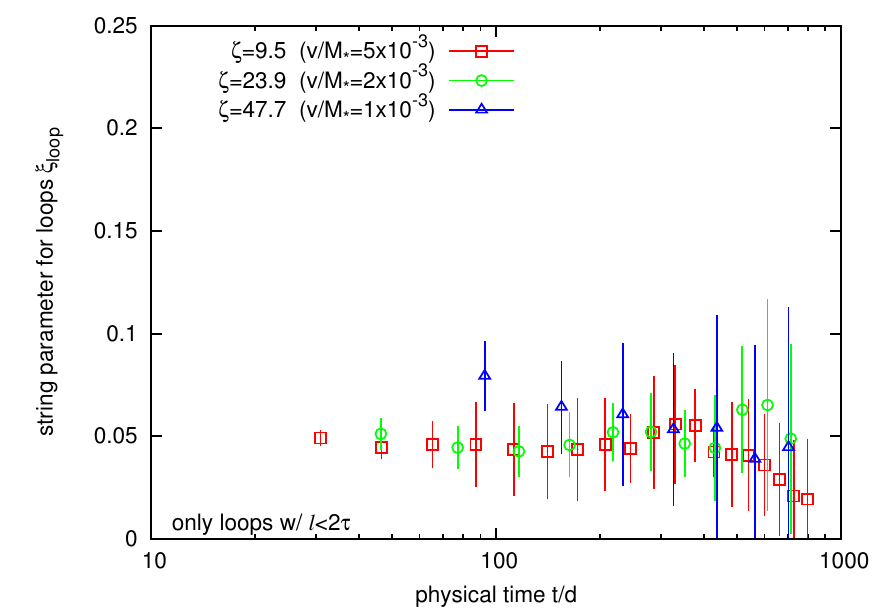} &
\includegraphics[scale=.8]{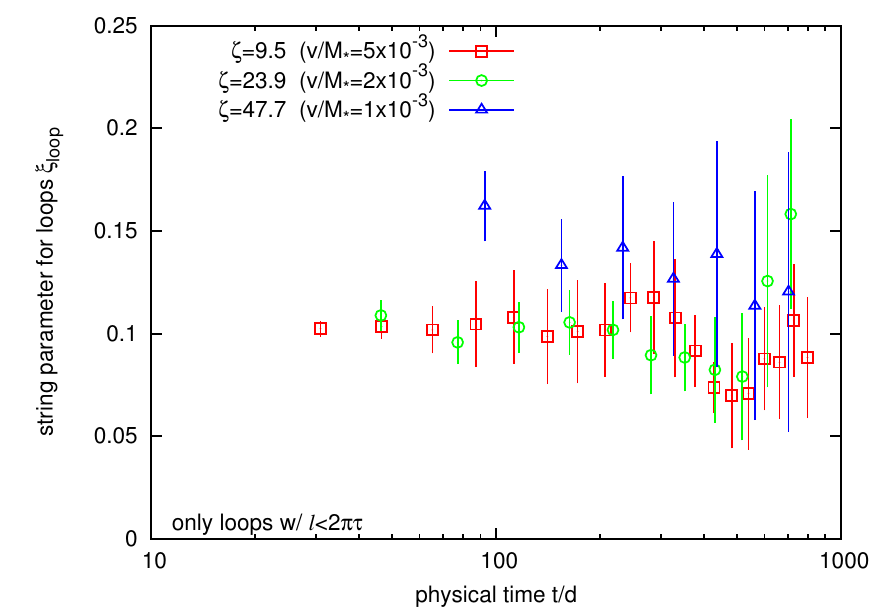}
\end{tabular}
\caption{\label{fig:loop}
Evolution of string parameter contributed only from the loops with physical circumference length less than $t$ (left) and $\pi t$ (right).
Color and symbol are the same as in Figure \ref{fig:xi}.
}
\end{figure}

On the other hand, $\xi$ shown in Figure~\ref{fig:xi} incorporates both infinite strings and loops. Thus, one may wonder
the logarithmic growth occurs only among loops or infinite strings. 
To answer this question, in Figure~\ref{fig:loop} we plot the string parameter only contributed from loops with 
length less than $t$ (left) and $\pi t$ (right). From the figure, one can see that the contribution of the loops 
in the total length is around 10 percent.  Thus, the dominant contribution in $\xi$ comes from the infinite strings.
In addition, the figure shows that the string parameter of loops does not apparently grow in time. Therefore, we conclude that 
the logarithmic growth seen in Figure~\ref{fig:xi} originates from the infinite strings.

\begin{figure}
\centering
\includegraphics[scale=.8]{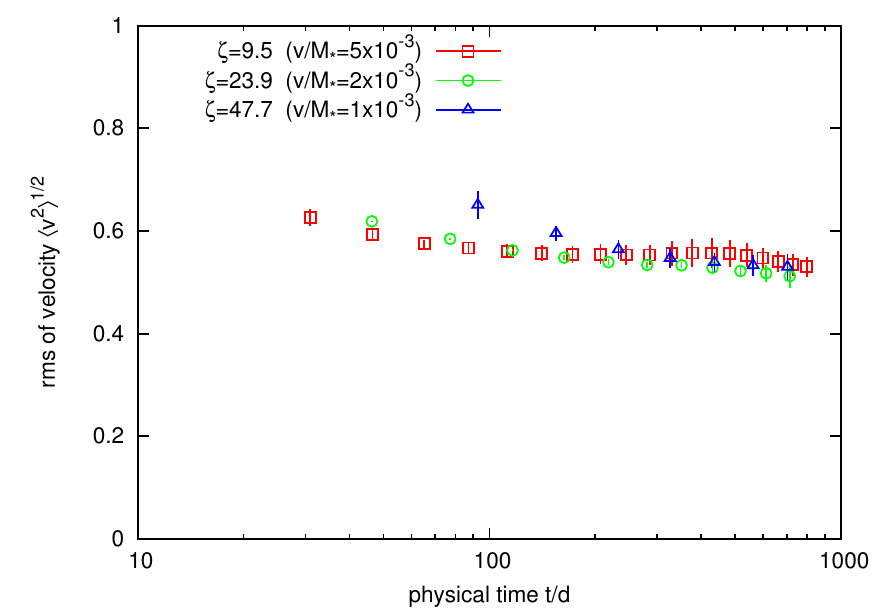} 
\caption{\label{fig:velocity}
Evolution of the rms of velocity. Color and symbol are the same as in Figure~\ref{fig:xi}.
}
\end{figure}

Velocity of strings also exhibits some degree of discrepancy from the previous studies.
The root mean square of string velocity $\sqrt{\langle \bm v^2\rangle}$ is plotted in Figure~\ref{fig:velocity}. 
We found $0.6\lesssim\sqrt{\langle \bm v^2\rangle}\lesssim 0.7$ around the initial stage, which is consistent with \cite{Yamaguchi:2002sh}.
However, as time advances, it decreases gradually and reads around 0.5 at the later time of the simulation. 
This is substantially smaller compared to the velocity estimated from simulation of local-string based 
on the Nambu-Goto action \cite{Bennett:1989yp,Moore:2001px} but consistent with field-theoretic simulation of local strings \cite{Hindmarsh:2008dw}.
At this moment, we are unable to tell if the velocity has settled down already within the simulation time or continues to decrease subsequently.

\begin{figure}
\centering
\includegraphics[scale=.8]{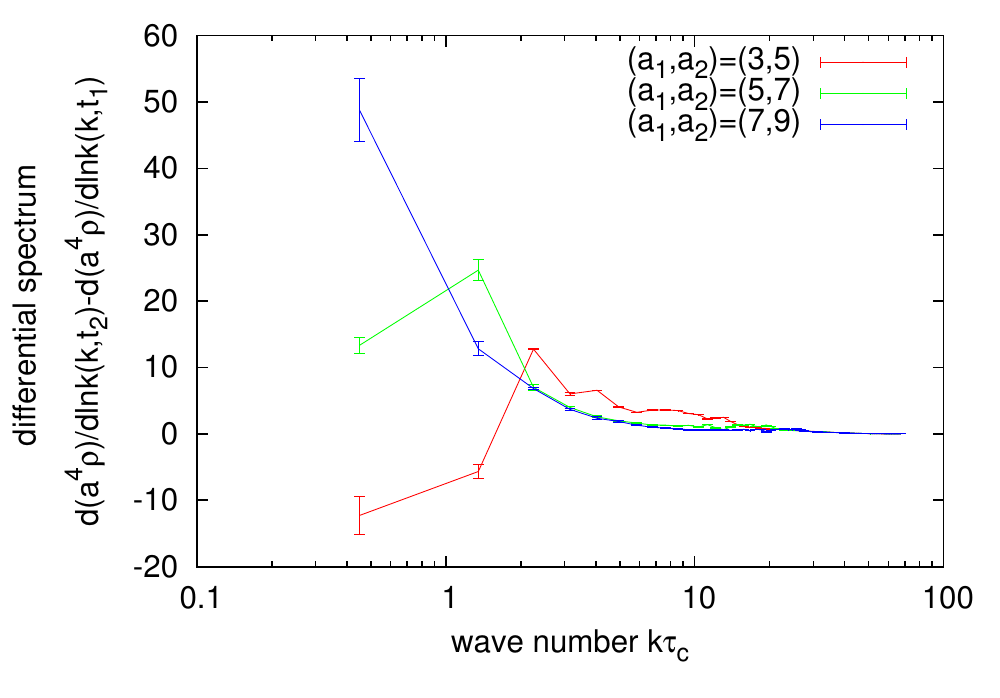} \\
\caption{\label{fig:diffspectrum}
Differential spectrum between two scale factors ($a_1$, $a_2$) for the case of $v/M_*=0.005$.
}
\end{figure}

In Figure \ref{fig:diffspectrum} we plot the evolution of the differential energy spectrum of the axion radiation,
$\frac{d(a^4\rho_{\rm rad})}{d\ln k}$, for $v/M_*=0.005$.
This demonstrates two aspects of the axion radiation from the axion strings. One is that energy spectrum has a peak
at low momenta and decays quickly towards higher ones. The other is that the spectral peaks moves towards lower momenta
as time advances. 

To quantify the evolution of the spectral peaks, let us define 
\be
\epsilon(t)= \frac t{2\pi a \overline{k^{-1}}(t)},
\label{eq:epsilon}
\ee
where $\overline{k^{-1}}$ is the mean inverse momentum of radiated axion at $t$ defined as 
\be
\overline{k^{-1}} \equiv 
\frac{\frac d{dt}\int dk k^{-1}P_{\rm rad}(k,t)}
{\frac d{dt} \int dk P_{\rm rad}(k,t)}
\ee
with $a^{-4}\int dk P_{\rm rad}$ being the energy density of the axion radiation.
Roughly speaking, $\epsilon$ represents the mean wave number of radiated axion in units of the Hubble expansion rate.

\begin{figure}
\centering
\includegraphics[scale=.8]{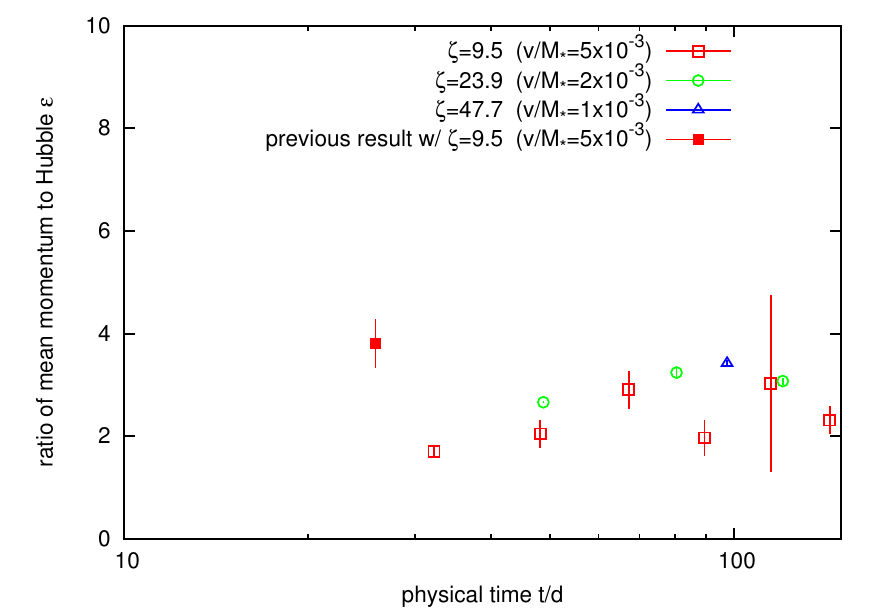} \\
\caption{\label{fig:eps}
Evolution of the mean momentum of radiated axions in units of the Hubble expansion rate (see Eq.~\eqref{eq:epsilon}).
Color and symbol are the same as in Figure \ref{fig:xi}.
}
\end{figure}

Figure \ref{fig:eps} shows the evolution of $\epsilon$. 
We confirm $\epsilon$ lies between two and four, 
which is largely consistent with \cite{Yamaguchi:1998gx,Hiramatsu:2010yu}.
This supports the claim that radiated axions have typical momenta comparable to the Hubble expansion rate 
\cite{Davis:1985pt,Davis:1986xc,Davis:1989nj,Battye:1993jv,Battye:1994au}.
Due to the limitation of the number of bins, which is taken to 75 in our analysis, 
we cannot resolve the spectral peaks any longer when the mean wave number becomes comparable to the bin size.
This is why we can trace $\epsilon$ only up to $t/d\sim 100$.

On the other hand, the wavelength of radiated axions are expected to be related 
to the correlation length of the strings, which is proportional to $1/\sqrt{\xi}$. This implies $\epsilon$ should be 
proportional to $\sqrt{\xi}$. Nonetheless, at this moment we cannot confirm such evolution of $\epsilon$ 
due to the lack of late-time measurements of $\epsilon$. 

\section{Discussion} 
\label{sec:Discussion}
Among the results we presented in the previous section, 
the growth of string parameter would have the most profound significance.
While field theoretic simulation of physical strings cannot trace $t/d\gtrsim\mathcal(N_g^{1/3})$, 
$\log(t/d)$ can be as large as $70$ at the time of the QCD phase transition 
given $10^{9}\,{\rm GeV} \lesssim v\lesssim10^{11}$\,GeV,
where the lower and upper bounds come from astrophysical consideration and cosmological axion abundance, 
respectively~\cite{Turner:1989vc,Raffelt:1990yz,Tanabashi:2018???}.
This means even the merely logarithmic growth can enhance $\xi$ by 
a factor of seven or larger \footnote{
Fitting $\xi$ with a power-law dependence on $\log(t/d)$ at $\log(t/d)\ge 300$, 
the enhancement factor can be $4.6\pm 0.68$, $8.8\pm 1.6$ and $3.8\pm 0.4$ for 
$v/M_*=0.005$, 0.002, and 0.001, respectively.
}. 

The number of axions radiated from strings is proportional to the number of axion strings
and inversely proportional to the typical momentum of radiated axions.
The former is proportional to $\xi$. On the other hand, the latter is expected to be proportional to $\sqrt{\xi}$, 
though we couldn't confirm this in our present simulation.
Assuming that the axion momentum grows in proportion to $\sqrt{\xi}$, we expect the axion abundance is proportional $\sqrt{\xi}$.
Our results indicates the abundance of axion can be enhanced by a factor of two or three.
This should give a large impact not only on the constraints
on the decay constant $f_a$ from overabundance of the axion CDM but also detection experiments, 
which target the mass of axion or $f_a\propto v$ with which the axion accounts for the observed CDM density.

Logarithmic growth in the string parameter is reported in
\cite{Fleury:2015aca,Klaer:2017qhr,Klaer:2017ond}.  However, the
relation between our observation and theirs is not at all clear.  This
is because there is a crucial difference between the two simulations.
Our strings are physical ({\it i.e.} constant string width), while the strings
in \cite{Fleury:2015aca,Klaer:2017qhr,Klaer:2017ond} are so-called fat
strings ({\it i.e.} string width growing in time). As discussed in
\cite{Dabholkar:1989ju}, the radiation backreaction, which should be
crucial in the dynamics of global strings
\cite{Davis:1985pt,Vilenkin:2000jqa}, is suppressed by the logarithm of
$t/d$ (see also \cite{Moore:2001px}). 
Since $d$ increases in fat strings, the extent of suppression of the radiation backreaction
may deviate from that in physical strings. 
Therefore, results of fat strings cannot be easily identified with physical ones.

What is more puzzling is the nontrivial dependence of $\xi$ on the PQ breaking scale $v$.
As seen in Figure~\ref{fig:xi}, we found the magnitude of $\xi$ depends on $v$.
Although previously reported values of $\xi$ have also been varying among literature 
\cite{Yamaguchi:1998gx,Moore:2001px,Yamaguchi:2002sh,Hiramatsu:2010yu},
the discrepancy has not been paid much attention due to statistical errors. 
Thanks to larger box size available at given $t$, the discrepancy in $\xi$ is manifested 
in our analysis with enough significance. 

Such dependence apparently conflicts with the prediction of one-scale model \cite{Kibble:1984hp,Bennett:1985qt,Martins:2000cs}. 
Moreover, the dependence of $\xi$ on $v$ is not monotonic. 
As can be seen from Figure~\ref{fig:xi}, the magnitude of $\xi$ fluctuates as $v/M_*$ is 
decreased from $0.005$ to $0.001$. This may indicate existence of competing effects, 
which explicitly depend on the PQ breaking scale. 

At the moment, the physical origin of the logarithmic growth of the string parameter $\xi$ and 
its dependence on $v$ is still cloaked. 
As noted above, the efficiency of the energy loss of strings through radiation is suppressed by $\log(t/d)$. 
We speculate this can at least partially account for the logarithmic growth of $\xi$.
Nonetheless, more rigorous investigation is required before we conclude this is the dominant cause.
In view of the impact, scrutiny of the origin cannot be stressed too much.
We pursue more detailed analysis with extensive parameter range and physical interpretation 
in the future publication \cite{Kawasaki}.

We have also computed the spectrum of the axion radiated from the axion strings. 
The result is largely in agreement with the previous studies \cite{Yamaguchi:1998gx,Hiramatsu:2010yu}.
We have confirmed the spectral peak stays around a few time larger than the Hubble expansion rate.

\bigskip
{\it Note added:} 
After we have finished writing the substantial part of the present manuscript
we became aware of a paper by Gorghetto, Hardy, and
Villadoro, who performed very similar numerical analysis
\cite{Gorghetto:2018myk}.

\section*{Acknowledgements}
This research is supported by JSPS KAKENHI Grant Numbers 17H01131 (M.K.), 17K05434 (M.K.), 
JP25287054 (M.Y.), JP18H04579 (M.Y.), JP15H02082 (J.Y. and T.S.), 18H04339 (T.S.), 18K03640 (T.S.), 
Grant on Innovative Areas JP15H05888 (M.Y. and J.Y.) and MEXT KAKENHI Grant Number 15H05889 (M.K.).
This work is also supported by WPI Initiative, MEXT, Japan (M.K.).
This research used computational resources of COMA and Oakforest-PACS provided by Interdisciplinary 
Computational Science Program in Center for Computational Sciences, University of Tsukuba. 



\end{document}